\documentclass[prl,reprint,amsmath,amssymb]{revtex4-2}
\usepackage{times}
\usepackage{graphicx}
\usepackage{epsfig}
\usepackage{dcolumn}
\usepackage{bm}
\usepackage{color}
\usepackage{longtable}
\usepackage{array}
\usepackage{multirow}
\usepackage{overpic}
\usepackage{lineno}
\usepackage[breaklinks, colorlinks, linkcolor=blue, anchorcolor=blue, citecolor=blue, urlcolor=blue]{hyperref}

\begin{document}

\title{Evidence for the Cusp Effect in $\eta'$ Decays into $\eta\pi^0\pi^0$}
\author{
\begin{small}
\begin{center}
M.~Ablikim$^{1}$, M.~N.~Achasov$^{11,b}$, P.~Adlarson$^{70}$, M.~Albrecht$^{4}$, R.~Aliberti$^{31}$, A.~Amoroso$^{69A,69C}$, M.~R.~An$^{35}$, Q.~An$^{66,53}$, X.~H.~Bai$^{61}$, Y.~Bai$^{52}$
% ...
\\
\vspace{0.2cm}
(BESIII Collaboration)\\
\vspace{0.2cm} {\it
$^{1}$ Institute of High Energy Physics, Beijing 100049, People's Republic of China\\
% ...
$^{4}$ Bochum Ruhr-University, D-44780 Bochum, Germany\\
% ...
$^{11}$ G.I. Budker Institute of Nuclear Physics SB RAS (BINP), Novosibirsk 630090, Russia\\
% ...
$^{31}$ Johannes Gutenberg University of Mainz, Johann-Joachim-Becher-Weg 45, D-55099 Mainz, Germany\\
% ...
$^{35}$ Liaoning Normal University, Dalian 116029, People's Republic of China\\
% ...
$^{52}$ Southeast University, Nanjing 211100, People's Republic of China\\
$^{53}$ State Key Laboratory of Particle Detection and Electronics, Beijing 100049, Hefei 230026, People's Republic of China\\
% ...
$^{61}$ University of Jinan, Jinan 250022, People's Republic of China\\
% ...
$^{66}$ University of Science and Technology of China, Hefei 230026, People's Republic of China\\
% ...
$^{69}$ University of Turin and INFN, (A)University of Turin, I-10125, Turin, Italy; (B)University of Eastern Piedmont, I-15121, Alessandria, Italy; (C)INFN, I-10125, Turin, Italy\\
$^{70}$ Uppsala University, Box 516, SE-75120 Uppsala, Sweden\\
% ...
$^{b}$ Also at the Novosibirsk State University, Novosibirsk, 630090, Russia\\
% ...
}
\end{center}
\vspace{0.4cm}
\end{small}
}

\date{February 26, 2023}

\begin{abstract}
Using a sample of $4.3\times 10^5$ $\eta'\to\eta\pi^0\pi^0$ events selected
from the ten billion $J/\psi$ event dataset collected with the BESIII detector,
we study the decay $\eta'\to\eta\pi^0\pi^0$ within the framework of
nonrelativistic effective field theory. Evidence for a structure at $\pi^+\pi^-$
mass threshold is observed in the invariant mass spectrum of $\pi^0\pi^0$ with
a statistical significance of around $3.5\sigma$, which is consistent with the
cusp effect as predicted by the nonrelativistic effective field theory. After
introducing the amplitude for describing the cusp effect, the $\pi\pi$
scattering length combination $a_0-a_2$ is determined to be
$\rm 0.226\pm0.060_{stat}\pm0.013_{syst}$, which is in good agreement with
theoretical calculation of $0.2644\pm0.0051$.
\end{abstract}

\maketitle

\begin{figure}[!htbp]
  \centering
  \includegraphics[width=0.9\linewidth]{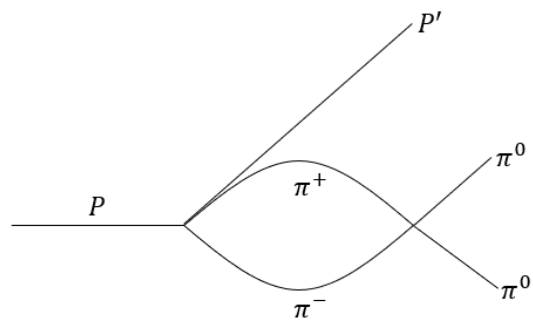}
  \caption{
            One-loop contribution in $P\to P'\pi^0\pi^0$ decay, where $P$ and
            $P'$ denote pseudoscalar particles in initial and final states, respectively.
            Different behaviors below and above the charged pion mass
            threshold cause the cusp effect.
          }
  \label{topo}
\end{figure}

Experimental studies of light meson decays are important guides to our
understanding of how QCD works in the nonperturbative regime.
In this contest, the $\pi\pi$ and $\pi K$ interactions at low energies have been the
subject of investigations for a few decades. In $\pi\pi$ interaction, one of the prominent features
is the loop contribution to the $\pi\pi$ scattering: the
$S$-wave charge-exchange rescattering $\pi^+\pi^-\to\pi^0\pi^0$ (as shown in
Fig.~\ref{topo}) causes a prominent cusp at the center of mass energy
corresponding to the summed mass of two charged pions. The cusp effect can shed
light on the fundamental properties of QCD at low energies, by determining the
strength of the $S$-wave $\pi\pi$ interaction~\cite{Meissner:1997fa,Cabibbo:2004gq,Cabibbo:2005ez,Colangelo:2006va,Gamiz:2006km,Bissegger:2008ff}.
Six decades ago this effect was predicted to be seen in $K^+\to\pi^0\pi^0\pi^+$~\cite{Budini:1961bac},
and it was finally observed in 2006~\cite{NA482:2005wht} by the NA48/2 experiment and studied  
further~\cite{Batley:2009ubw}. These results inspired theoretical predictions for the
cusp in other decays, such as $K_L\to 3\pi^0$~\cite{Cabibbo:2005ez,Gamiz:2006km,Bissegger:2007yq}
and $\eta\to 3\pi^0$~\cite{Ditsche:2008cq,Gullstrom:2008sy},
which were experimentally investigated:
it was observed in the decay of $K_L\to 3\pi^0$
by KTeV~\cite{KTeV:2008gel}, while no clear evidence was seen in
$\eta\to 3\pi^0$ decay~\cite{WASA-at-COSY:2008rsh,CrystalBallatMAMI:2008cye,
CrystalBallatMAMI:2008pqf,A2:2018pjo}.

Another process where the cusp effect is expected to have a sizable contribution is the hadronic decay $\eta'\to\eta\pi^0\pi^0$~\cite{Kubis:2009sb};
this has been experimentally investigated by BESIII~\cite{BESIII:2017djm}, with 5.6$\times$10$^4$ $\eta'\to\eta\pi^0\pi^0$ events, and no evidence was seen,
while the A2 experiment~\cite{Adlarson:2017wlz} accumulated about 1.24$\times$10$^5$ $\eta'\to\eta\pi^0\pi^0$ decays and reported a deviation
with a significance of about $2.5\sigma$ and it is also studied in chiral perturbation theory~\cite{Gonzalez-Solis:2018xnw}. Therefore, it is essential
to further investigate this decay with higher precision.

The recently available data of ten billion $J/\psi$ events~\cite{BESIII:2021cxx}
at BESIII imply an increased data sample of $\eta^\prime$ decays by nearly an
order of magnitude, offering a unique opportunity for further investigations of
the cusp effect. In this Letter, we present the
first evidence of the cusp effect in $\eta'\to\eta\pi^0\pi^0$ and the 
corresponding measurement of the $\pi\pi$ scattering length based on the 
nonrelativistic effective field theory (NREFT)~\cite{Kubis:2009sb}.

The BESIII detector~\cite{BESIII:2009fln,BESIII:2020nme}
records symmetric $e^+e^-$ collisions
provided by the BEPCII storage ring~\cite{Yu:2016cof}.
The cylindrical core of the BESIII detector
covers 93\% of the full solid angle and consists of a helium-based multilayer
drift chamber~(MDC), a plastic scintillator time-of-flight~(TOF) system, and a
CsI(Tl) electromagnetic calorimeter~(EMC), which are all enclosed in a
superconducting solenoidal magnet providing a 1.0~T (0.9~T in 2012) magnetic
field. The end cap TOF system was upgraded in 2015 using multigap resistive
plate chamber technology~\cite{MRPC:1,MRPC:2,Cao:2020ibk}.

To reconstruct events of $J/\psi\to\gamma\eta'$ with $\eta'\to\eta\pi^0\pi^0$,
the $\pi^0$ and $\eta$ are selected by $\pi^0/\eta\to2\gamma$ process. The
charged tracks are reconstructed from hits in the MDC. The polar angle with respect to 
the MDC symmetry axis should be in the range $|\cos\theta|<0.93$. The distance away
from the interaction point should be less than 10.0 cm in the beam direction and
1.0 cm in the radial direction. The photon candidates are reconstructed using
clusters of energy deposited in the EMC. The energy deposited in the nearby TOF system
is included in EMC measurements to improve the reconstruction efficiency and
the energy resolution. Photon candidates are required to have a deposited
energy larger than 25 MeV in the barrel region ($|\cos\theta|<0.80$) and 50 MeV
in the end cap regions ($0.86<|\cos\theta|<0.92$). A requirement on the EMC
cluster timing with respect to the most energetic photon, $-500<T<500$ ns, is
used to suppress electronic noise and energy deposits unrelated to the event.
The events with at least seven photon candidates and no charged tracks are
kept for further analysis.

\begin{figure}[!htbp]
  \centering
  \includegraphics[width=0.9\linewidth]{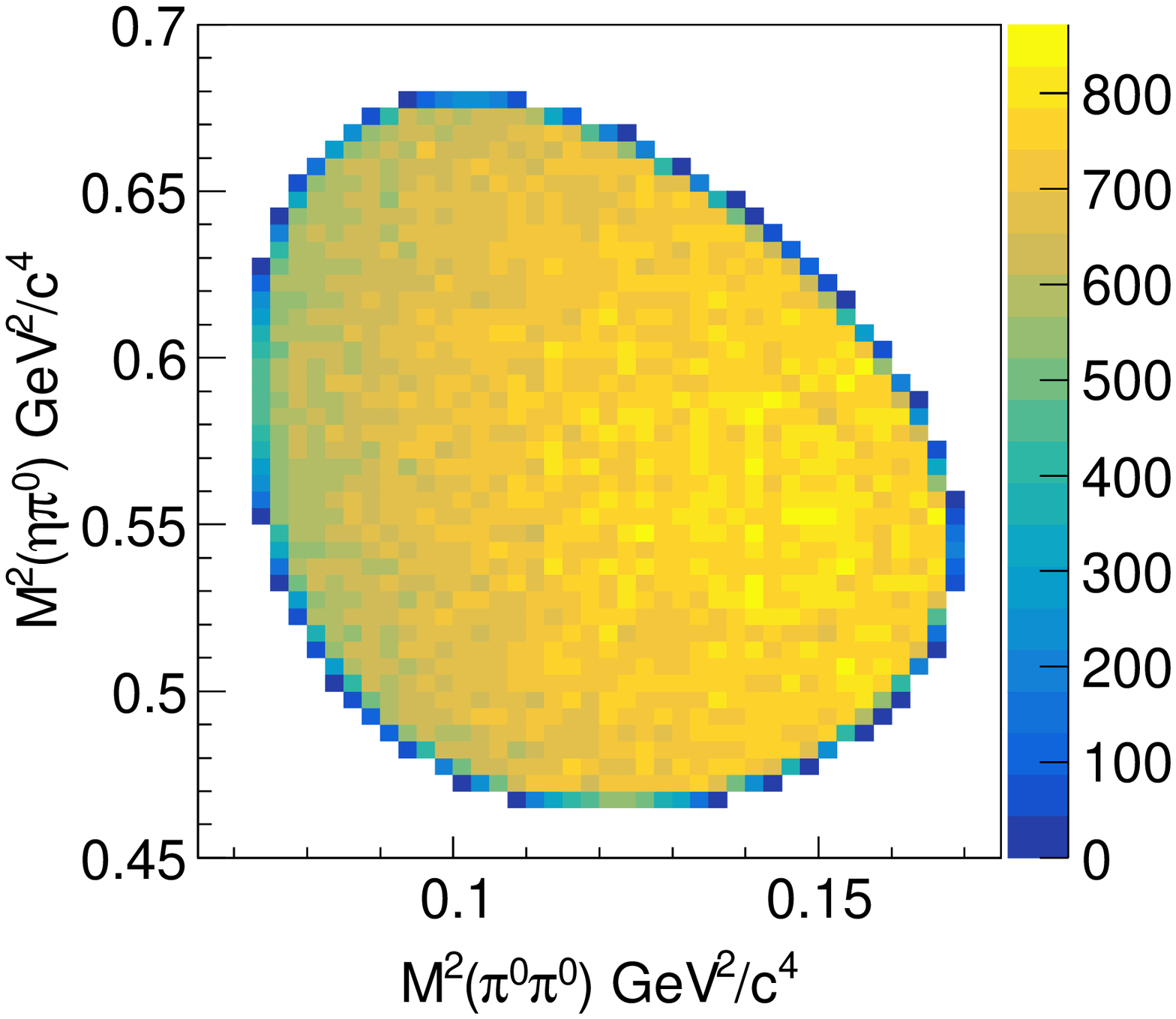}
  \caption{
            Dalitz plot of $\eta'\to\eta\pi^0\pi^0$.
          }
  \label{dalitz}
\end{figure}

For each candidate event, the photon with the maximum energy is assumed to be
the radiative photon originating from the decay of $J/\psi$, while the
remaining photons are used to reconstruct $\pi^0/\eta$ candidates. A
one-constraint (1C) kinematic fit is performed by constraining the invariant mass of photon
pairs to the $\pi^0$ or $\eta$ mass, and the $\chi^2$ for this fit
is required to be less than 25. Since the $\pi^0$ decays into two photons
isotropically in its rest frame, the angle of one photon in the $\pi^0$ rest
frame with respect to the $\pi^0$ momentum direction is required to satisfy
$|\cos\theta_{\pi^{0}}|<0.95$. Afterward, an eight-constraint (8C)
kinematic fit is performed for the $\gamma\eta\pi^0\pi^0$ combinations, requiring
energy-momentum conservation and constraining the invariant masses of the three
photon pairs to the nominal $\pi^0/\eta$ masses and of the $\eta\pi^0\pi^0$
combinations to the $\eta'$ mass. If more than one $\gamma\eta\pi^0\pi^0$
combination is found, only the one with the least $\rm \chi^{2}_{8C}$ is
retained. After the requirement of $\rm \chi^2_{8C}<100$, 432295 candidate events
are accepted for further analysis; the corresponding Dalitz plot is shown in
Fig.~\ref{dalitz}.

\begin{figure}[!htbp]
  \centering
  \includegraphics[width=0.8\linewidth]{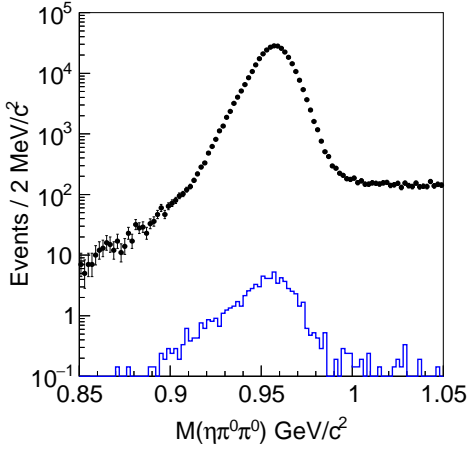}
  \caption{
           The $\eta\pi^{0}\pi^{0}$ invariant mass distribution  
           after 6C kinematic fit without the $\eta$ and $\eta'$ mass
           constraints. The dots with error bars are experimental data,
           and the blue histogram is the $\eta'\to3\pi^{0}$ peaking
           background from MC sample.
          }
  \label{bkg}
\end{figure}

To investigate the background contamination, a 6C kinematic fit, instead of the
8C fit, is performed on candidate events, in which the constraints on the masses of $\eta$ and $\eta'$ are
removed.
Figure~\ref{bkg} shows the $\eta\pi^0\pi^0$ invariant mass distribution of the data sample, after
requiring $\rm \chi^2_{6C}<100$ and $\eta$ mass window cut
$|M(\gamma\gamma)-M_{\eta}|<30$ MeV/$c^{2}$ on the unconstrained photon pair,
a clear $\eta'$ peak is observed.
In addition, a ten billion $J/\psi$
inclusive decay Monte Carlo (MC) sample generated with 
{\sc LUNDCHARM}~\cite{Chen:2000tv,Yang:2014vra} is used to check possible
background sources; the surviving events mainly
consist of the peaking background $\eta'\to3\pi^0$ decay channel and the flat contribution
from $J/\psi\to\omega\eta$, with $\omega\to\gamma\pi^0$ and $\eta\to3\pi^0$.
The background contamination rate is estimated about 0.82\%, and its shape on
$\pi^0\pi^0$ and $\eta\pi^0$ mass spectrum is smooth; therefore, it is neglected in the further analysis.

Using an unbinned maximum likelihood method, we fit the Dalitz plot of
$M^2(\pi^0\pi^0)$ versus $M^2(\eta\pi^0)$ within the framework of NREFT. 
(More details are given in the Supplemental Material~\cite{supp}.) The resolution effect and
detection efficiency are studied by MC simulation and taken into account in the fit.

\begin{figure}[!bthp]
  \centering
  \begin{overpic}[width=0.495\linewidth]{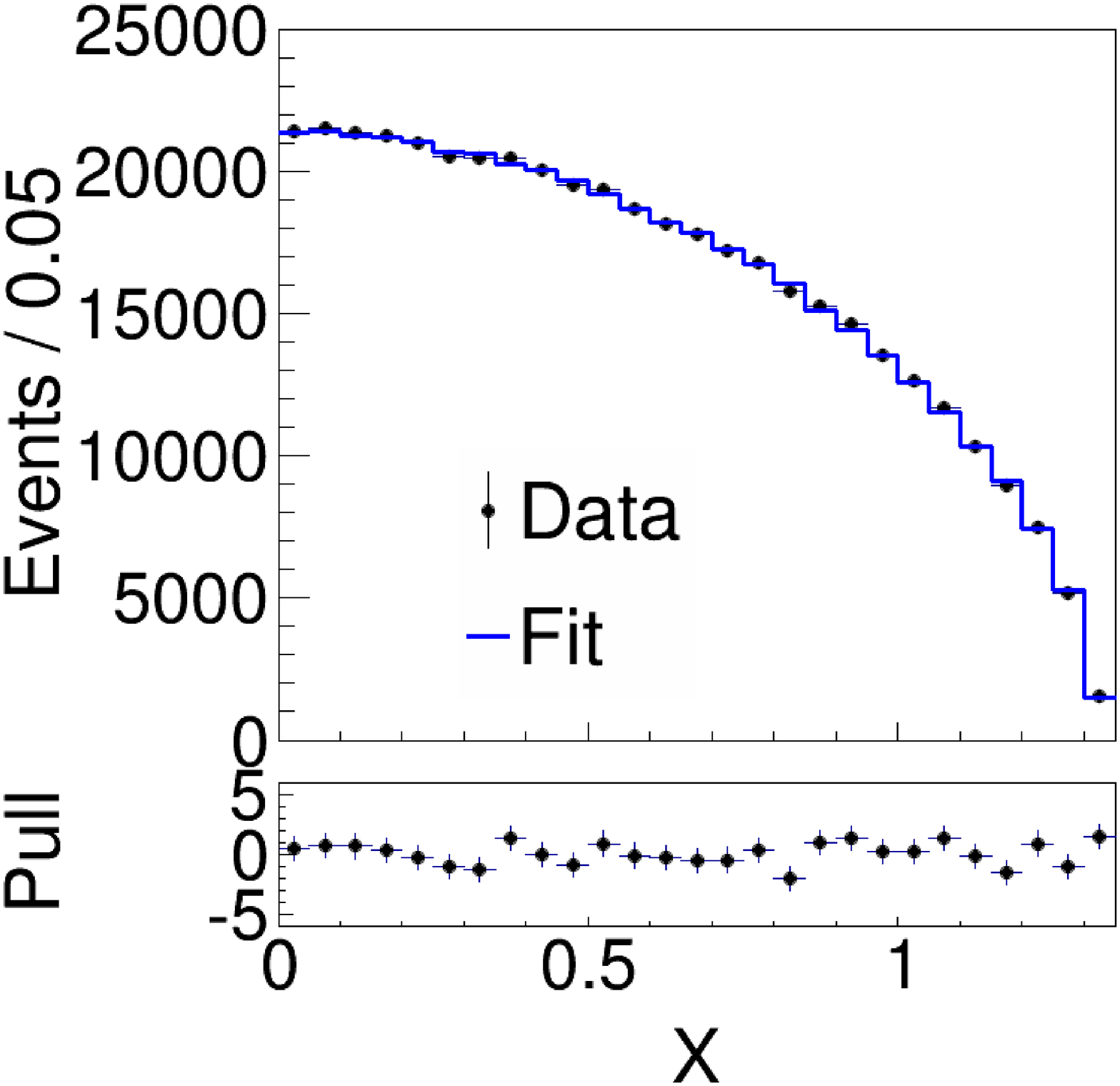}
    \put(80,85){(a)}
  \end{overpic}
  \begin{overpic}[width=0.495\linewidth]{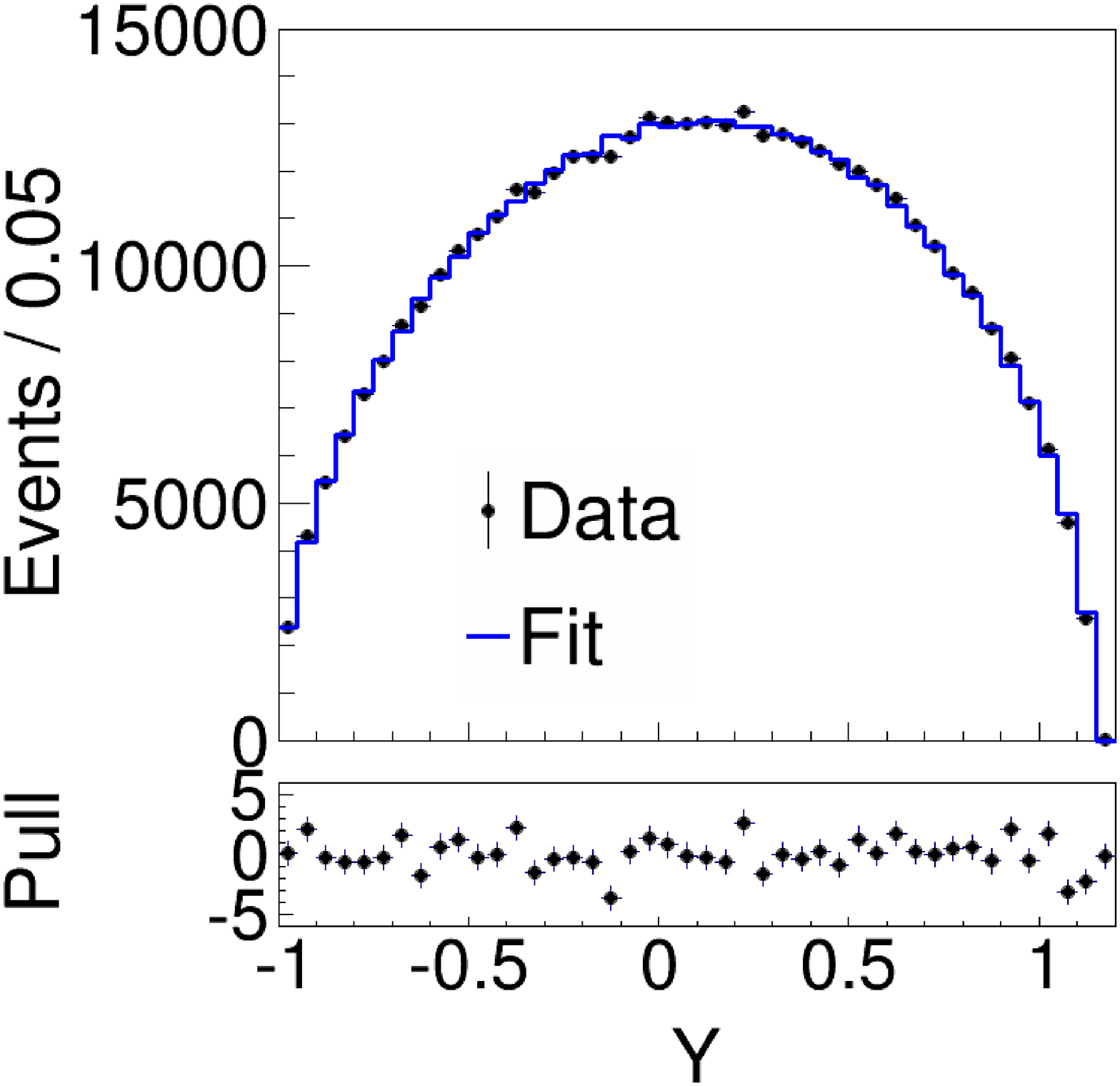}
    \put(80,85){(b)}
  \end{overpic}
  \begin{overpic}[width=0.495\linewidth]{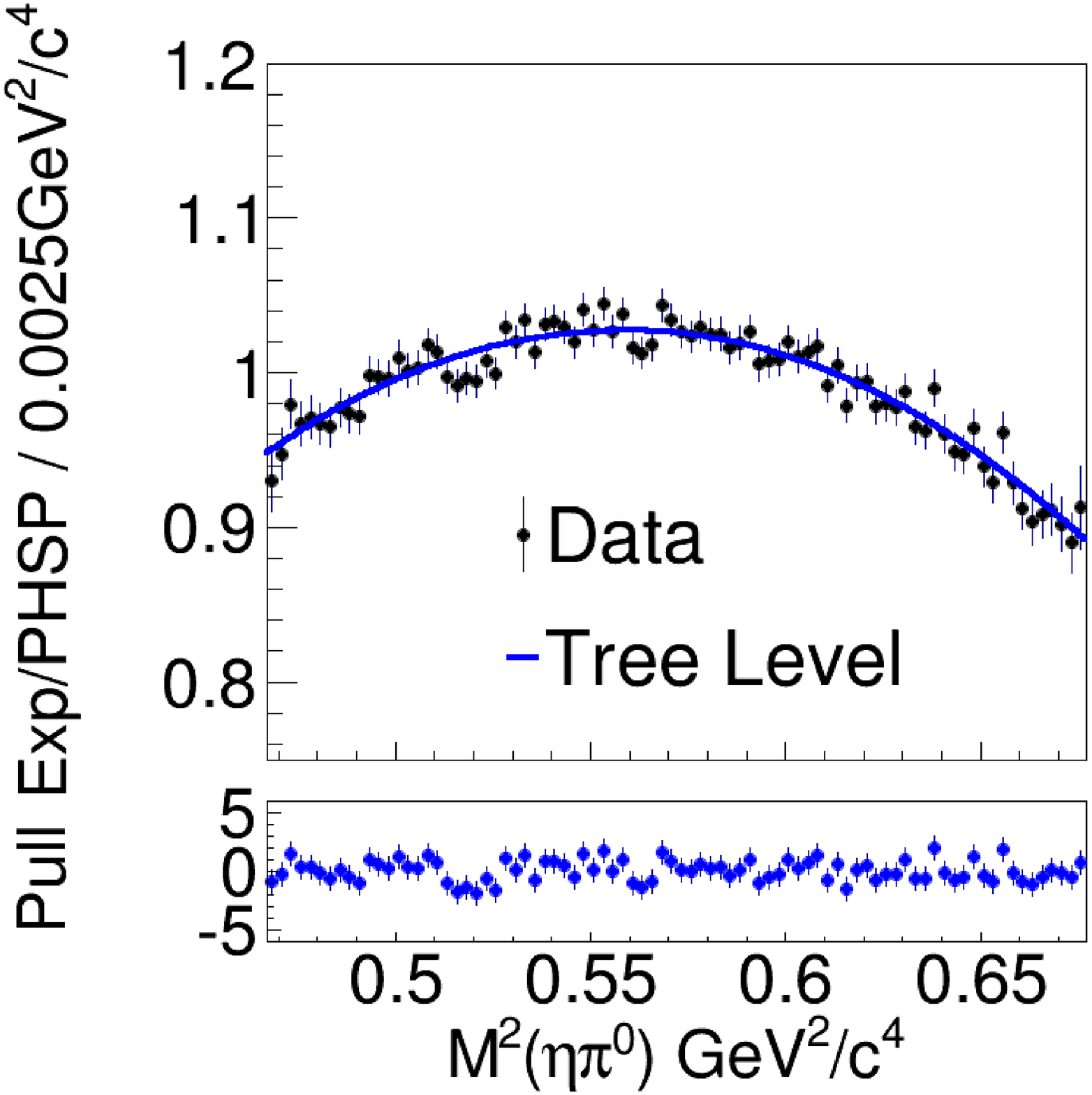}
    \put(80,85){(c)}
  \end{overpic}
  \begin{overpic}[width=0.495\linewidth]{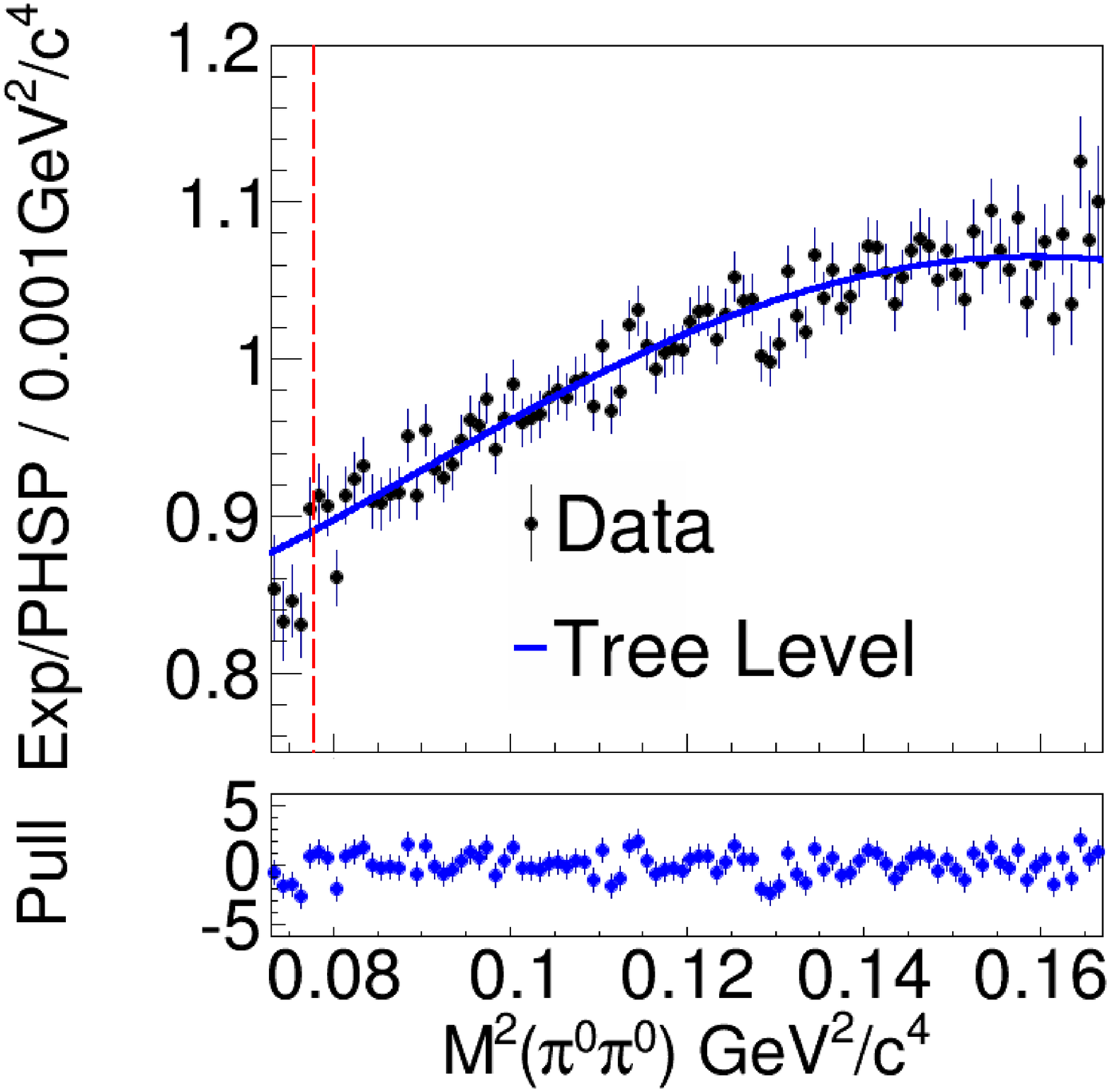}
    \put(80,85){(d)}
  \end{overpic}
  \caption{
            The fit result of fit I. The projections to $X$ and $Y$ are
            shown in (a) and (b), and the mass spectrums of
            $M^{2}(\eta\pi^{0})$ and $M^{2}(\pi^{0}\pi^{0})$ divided by phase
            space are shown in (c) and (d). The black dots with error bars are
            from data and the blue line is the fit result of the tree level amplitude.
            The red dashed line indicates the charged pion mass threshold.
          }
  \label{tree}
\end{figure}

\begin{table*}[htpb]
\centering
\caption{Experimental values of the matrix element parameters for $\eta'\to\eta\pi^{0}\pi^{0}$.}
\label{table_sum}
\begin{tabular}{ccccc}
      \hline
      \hline
      Parameters                  & Fit I                      & Fit II           & Fit III          & Fit IV                   \\
      \hline
      $a$                         & $-0.075\pm0.003\pm 0.001$  & $-0.207\pm0.013$ & $-0.143\pm0.010$ &$-0.077\pm0.003\pm 0.001$ \\
      $b$                         & $-0.073\pm0.005\pm 0.001 $ & $-0.051\pm0.014$ & $-0.038\pm0.006$ &$-0.066\pm0.006\pm 0.001$ \\
      $d$                         & $-0.066\pm0.003\pm0.001$   & $-0.068\pm0.004$ & $-0.067\pm0.003$ &$-0.068\pm0.004\pm 0.001$ \\ 
      $a_0-a_2$                   & -                          & $0.174\pm0.066$  & $0.225\pm0.062$  &$0.226\pm0.060\pm 0.013$  \\ 
      $a_0$                       & -                          & $0.497\pm0.094$  & -                & -                        \\
      $a_2$                       & -                          & $0.322\pm0.129$  & -                & -                        \\
      Statistical significance    & -                          & $3.4\sigma$      & $3.7\sigma$      & $3.6\sigma$              \\  
      \hline
      \hline
\end{tabular}
\end{table*}

In the simplest case (fit I), only the tree level contribution is
included in which the final state interaction effect is ignored. In this case, the amplitude
is the same as the general parametrization used in Ref.~\cite{BESIII:2017djm}.
The projections of the fit result to the Dalitz coordinates $X$ and $Y$ are 
shown in Figs.~\ref{tree}(a) and \ref{tree}(b), and they indicate a good description of data. 
The fitted parameter values, shown in Table~\ref{table_sum}, are
consistent with the previous BESIII measurement and the statistical
uncertainties are about one third of the previous results~\cite{BESIII:2017djm}.
In Figs.~\ref{tree}(c) and \ref{tree}(d) the comparisons between data and the fit projections
of the $\eta\pi^0$ and $\pi^0\pi^0$ invariant mass distributions
divided by the phase space are presented. The discrepancy between data and fit result below the
charged pion mass threshold corresponds to the cusp effect.
Therefore, we perform alternative fits by including the loop contributions
within the framework of NREFT to evaluate this effect, and fit I is
taken as the baseline for the further loop level fits.

At the loop level amplitude, only $\pi\pi$ scattering is considered while
$\eta\pi$ scattering is ignored; the $S$-wave $\pi\pi$ scattering lengths
$a_0$ and $a_2$ are included in the loop level amplitude by matching between
NREFT amplitude and partial wave decomposition,
\begin{equation}
\begin{split}
C_{00} &= \frac{16\pi}{3}(a_{0}+2a_{2})(1-\xi),          \\
C_{x}  &= \frac{16\pi}{3}(a_{2}-a_{0})(1+\frac{\xi}{3}), \\
C_{+-} &= \frac{8\pi}{3}(2a_{0}+a_{2})(1+\xi),          \\
\xi    &= \frac{M^{2}_{\pi^{\pm}}-M^{2}_{\pi^{0}}}{M^{2}_{\pi^{\pm}}}.
\end{split}
\end{equation}
where $C_{x}$ denotes the coupling coefficient of the cusp term
$\pi^+\pi^-\to\pi^0\pi^0$, and $C_{00}$ and $C_{+-}$ are the coupling
coefficients of noncusp terms $\pi^0\pi^0\to\pi^0\pi^0$ and
$\pi^+\pi^-\to\pi^+\pi^-$, which are defined in Ref.~\cite{Kubis:2009sb}.

The distribution of $M^2(\pi^0\pi^0)$ is determined by the whole amplitude and
all five parameters $a$, $b$, $d$, $a_0$, and $a_2$, while the distribution of
$M^2(\eta\pi^0)$ is mainly determined by parameter $d$, where $a$, $b$ and $d$
are coefficients in tree level amplitude.

To verify the prediction of NREFT and evaluate the scattering length
combination $a_0-a_2$, we perform many unbinned maximum likelihood fits in
different cases after including the contributions from the amplitudes at
one- and two-loop levels.

\begin{figure*}
  \centering
  \begin{overpic}[width=0.45\linewidth]{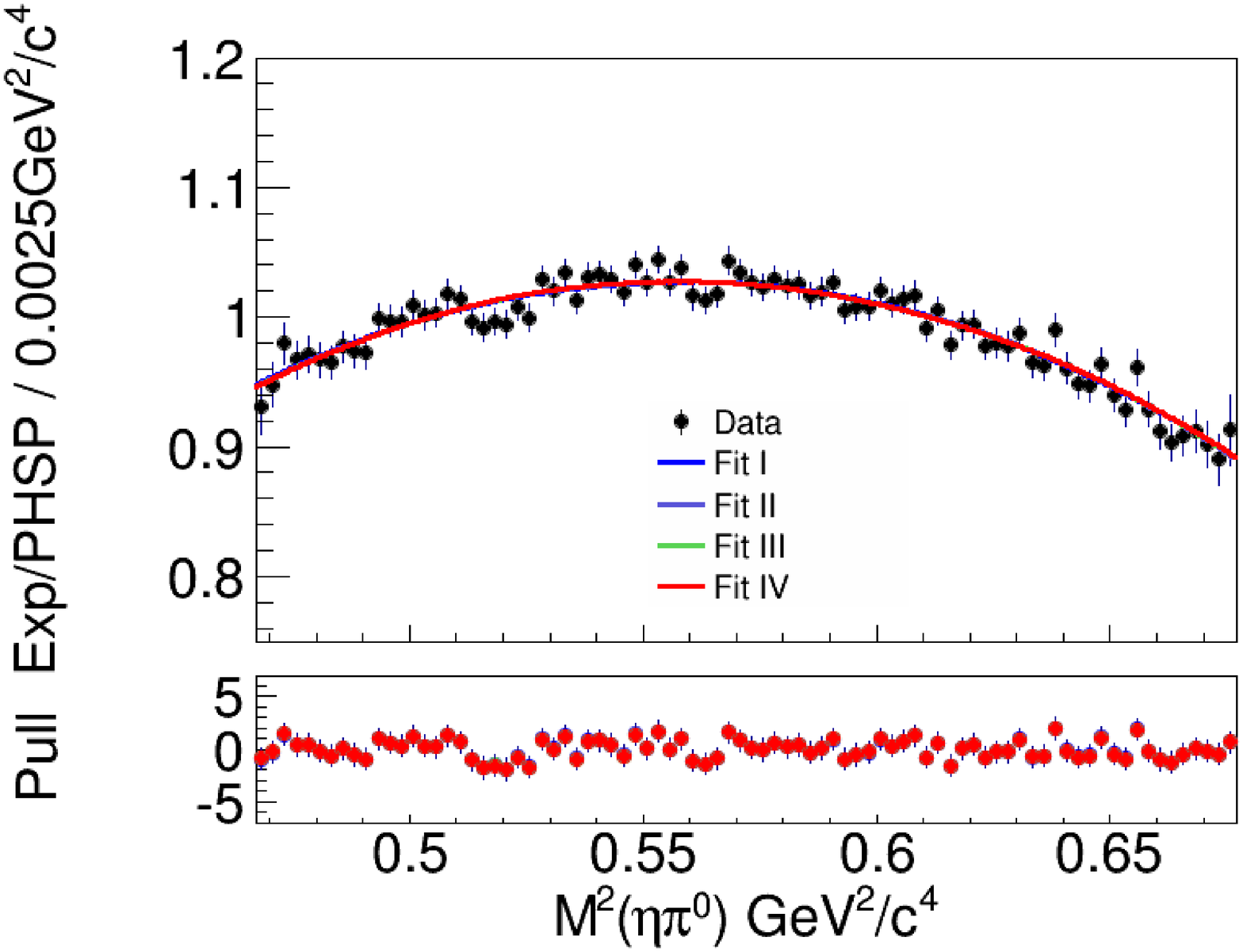}
    \put(80,65){(a)}
  \end{overpic}
  \begin{overpic}[width=0.45\linewidth]{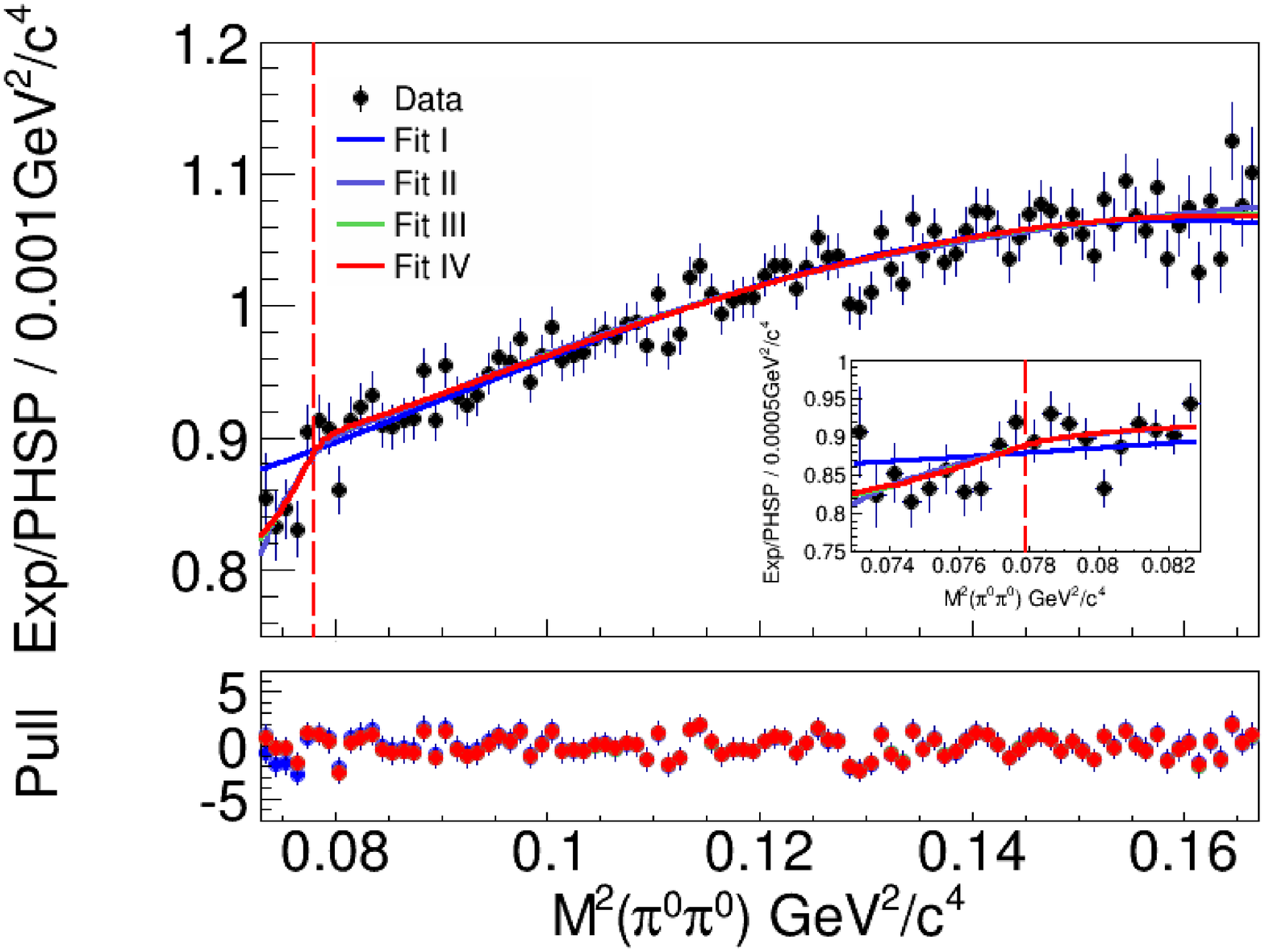}
    \put(80,65){(b)}
  \end{overpic}

  \caption{
            The fit result projections divided by phase space of different models to variable 
            (a)$M^{2}(\eta\pi^0)$ and (b)$M^{2}(\pi^0\pi^0)$. The black dots with error bars are
            from data. The solid lines are fit results from the corresponding models. The red
            dashed line indicates the charged pion mass threshold. The cusp region is also shown
            in the inset.
          }
  \label{loop}
\end{figure*}

In the case when all the parameters are free (fit II), the fit quality is 
improved, and we obtain a statistical significance of $3.4\sigma$ compared
to fit I. In Fig.~\ref{loop}, the comparison  between the fit and
data for the projections in different variables, as well as the pull distributions, shows
that the fit provides a good description, in particular, for the region below
the charged pion mass threshold. However, the correlation between the four
parameters $a$, $b$, $a_0$, and $a_2$ is very
large, as shown in Eq.~(\ref{fit2cor}). This strong correlation between $a$, $b$, $a_0$, and $a_2$ may be caused
by the loop level amplitude contribution to noncusp terms. The scattering
length combination is calculated to be $a_0-a_2=0.174\pm0.066$,

\begin{equation}
  \left(
  \begin{array}{c|cccc}
          &                 b &                 d &             a_{0} &             a_{2} \\ \hline
    a     & \hspace{7pt}0.831 & \hspace{7pt}0.189 &            -0.966 &            -0.789 \\
    b     &                   & \hspace{7pt}0.348 &            -0.918 &            -0.839 \\
    d     &                   &                   &            -0.257 &            -0.210 \\
    a_{0} &                   &                   &                   & \hspace{7pt}0.872 \\
  \end{array}
  \right)
  \label{fit2cor}
\end{equation}

To reduce the correlations between parameters, we also made an attempt
(fit III) by fixing $a_0+2a_2=0.1312$ according to
the theoretical values $a_0=0.220\pm0.005$ and $a_2=-0.0444\pm0.0010$~\cite{Kubis:2009sb}
and setting $a_0-a_2$ as a free parameter,
since only $C_x$ contributes to the cusp effect. The fit
result presented in Fig.~\ref{loop} shows a good agreement with
 data, also in the region below the charged pion mass
threshold. The fitted parameter values are summarized
in Table~\ref{table_sum}, and the corresponding correlations are shown in
Eq.~(\ref{fit3cor}); the obtained value of $a_0-a_2=0.225\pm0.062$ is in
agreement with the theoretical value $0.2644\pm0.0051$~\cite{Kubis:2009sb}. 
We also test by changing the value of $a_0+2a_2$ or fixing both $a_0+2a_2$
and $2a_0+a_2$ to theoretical values, the fit results are consistent with the
result of fit III, and $a_0-a_2$ is not sensitive to fixed value,

\begin{equation}
  \left(
  \begin{array}{c|ccc}
          &      b &                 d &     a_{0} - a_{2} \\ \hline
    a     & -0.560 &            -0.046 &            -0.955 \\
    b     &        & \hspace{7pt}0.249 & \hspace{7pt}0.457 \\
    d     &        &                   &            -0.032 \\
  \end{array}
  \right)
  \label{fit3cor}
\end{equation}

Comparing to the tree level amplitude, the loop contributions with $C_{00}$ and
$C_{+-}$ are expected to be small. Additionally, we performed an alternative 
fit (fit IV) by ignoring noncusp terms with $C_{00}$ and $C_{+-}$ and
only introducing the decay amplitude with $C_x$ for the description of the cusp
effect. In this case, the fitted values of
different parameters, summarized in Table~\ref{table_sum}, are in agreement with
those of fit I, the correlations shown in Eq.~(\ref{fit4cor}) are reduced
and the statistical significance of the cusp effect is
$3.6\sigma$, while the scattering length combination $a_0-a_2=0.226\pm0.060$ is
consistent with that in Ref.~\cite{Kubis:2009sb}. In addition, we found that
the log-likelihood value of fit IV is very close to those of fit II
fit III, which implies that the introduction of the loop contributions
with $C_{00}$ and $C_{+-}$ has little impact on the improvement of the fit
quality and the cusp effect, but significantly increases the correlations
between the different parameters. Therefore, in this analysis, it is reasonable to ignore 
these loop contributions in fitting data,

\begin{equation}
  \left(
  \begin{array}{c|ccc}
          &       b &             d     &     a_{0} - a_{2} \\ \hline
    a     &  -0.363 &            -0.253 & \hspace{7pt}0.126 \\
    b     &         & \hspace{7pt}0.257 & \hspace{7pt}0.237 \\
    d     &         &                   &            -0.107 \\
  \end{array}
  \right)
  \label{fit4cor}
\end{equation}

The systematic uncertainties for the Dalitz plots analysis are listed in 
Table~\ref{table5_1}. We calculate the total systematic uncertainty by 
assuming that all the contributions are independent and adding them in quadrature.

The photon detection efficiency is studied with the control sample of
$J/\psi\to\rho^0\pi^0$ events. To evaluate the impact from the slight
discrepancy between data and MC simulation, we perform a correction on
the photon detection and the change of the fit results is considered as the
systematic uncertainty.

To estimate the uncertainties from the 1C kinematic fit for $\pi^0$ and $\eta$,
we selected as control samples $J/\psi\to\pi^+\pi^-\pi^0$ and
$J/\psi\to\gamma\eta'$ with $\eta'\to\eta\pi^+\pi^-$, without kinematic fit. 
After taking into account the discrepancy between data and MC simulation,
repeating the fit with the weighted events leads to changes of the parameter
values, which are assigned as the systematic uncertainties.

To check if the photon miscombinations can effect the fitted parameters, we 
generate a MC sample based on NREFT amplitude and tag miscombination events by
matching the truth and the reconstructed value of photon momentum. Two fits
are performed to the sample with and without miscombination events, and the
change of the results is taken as the systematic uncertainty.

To evaluate the uncertainty associated with the efficiency parametrization,
we change the Dalitz plot variables to $M^2(\eta\pi)$ and $\cos\theta$, where
$\theta$ is the angle between the directions of the two $\pi$s in the rest frame of
$\eta\pi$. We repeat the fit based on the newly defined Dalitz plot variables, and
the change of the resulting parameters with respect to the nominal results is
assigned as the systematic uncertainty.

The uncertainty of the 8C kinematic fit mainly comes from the inconsistency of
the photon resolution between data and MC simulation. We adjust the energy
resolution in the reconstructed photon error matrix to ensure that the MC
simulation provides a good description of data. Afterward, an alternative
fit is performed and the change of the fitted parameters with respect to the
nominal result is taken as the systematic uncertainty.

To estimate the uncertainty from resolution effect, we vary the resolution by
$\pm10\%$ and perform alternative fits. The maximum change with respect to the
nominal result is taken as the systematic uncertainty.

\begin{table}
  \centering
  \caption{Summary of the systematic uncertainty sources and their corresponding contributions (\%).}
  \label{table5_1}
  \begin{tabular}{lccccccc}
    \hline
    \hline
    Parametrization                       & \multicolumn{3}{c}{ \bf Fit I}   & \multicolumn{4}{c}{\bf Fit IV}  \\
    \hline
    Source                                & $a$ & $b$ & $d$                  & $a$ & $b$ & $d$ & $a_{0}-a_{2}$ \\
    Photon detection                      & 0.7 & 0.4 & 1.0                  & 0.6 & 0.4 & 0.9 & 1.8           \\
    $\eta$ 1C kinematic fit               & 0.1 & 0.6 & 0.0                  & 0.1 & 0.7 & 0.0 & 0.2           \\
    $\pi^{0}$ 1C kinematic fit            & 0.1 & 0.2 & 1.0                  & 0.1 & 0.2 & 0.9 & 0.3           \\
    Photon miscombination                 & 0.0 & 0.2 & 1.1                  & 0.0 & 0.2 & 1.1 & 1.6           \\
    Efficiency presentation               & 0.7 & 1.0 & 0.4                  & 0.7 & 0.9 & 0.4 & 1.9           \\
    Kinematic fit                         & 0.5 & 1.3 & 0.7                  & 0.4 & 0.9 & 0.8 & 4.2           \\
    Resolution                            & 0.0 & 0.0 & 0.0                  & 0.1 & 0.3 & 0.0 & 2.0           \\
    Total                                 & 1.1 & 1.8 & 2.0                  & 1.0 & 1.6 & 1.9 & 5.6           \\
    \hline
    \hline
  \end{tabular}
\end{table}

In summary, using ten billion $J/\psi$ events collected with 
the BESIII detector, we select 
a $\eta'\to\eta\pi^0\pi^0$ sample 8 times larger
than that previously analyzed by BESIII, and perform a Dalitz plot analysis
within the framework of nonrelativistic effective field theory. The fit with tree level
amplitude shows a discrepancy below the charged pion mass threshold,
which implies the existence of the cusp effect. To describe the data in this region,
the contributions at one- and two-loop level are introduced in the decay
amplitude. We perform alternative analyses by taking into account the
cusp effect and the results are summarized in Table~\ref{table_sum}. For each
case, the amplitude provides a good description of the structure around the
charged pion mass threshold and the statistical significance is found to be
around $3.5\sigma$. The scattering length combination $a_0-a_2$ is measured to
be $0.226\pm0.060\pm 0.013$, which is in good agreement with the theoretical
value of $0.2644\pm0.0051$~\cite{Kubis:2009sb} within the uncertainties.
The observation of the evidence of the cusp effect in $\eta'\to\eta\pi^0\pi^0$
decay demonstrates the excellent potential to investigate the underlying
dynamics of light mesons at the BESIII experiment. The prospects~\cite{Fang:2021wes}
for the precise measurements are very promising at the planned Super Tau-Charm
Factories~\cite{Charm-TauFactory:2013cnj,Peng:2020orp}.

The BESIII Collaboration thanks the staff of BEPCII and the IHEP computing center for their strong
support. This work is supported in part by National Key R\&D Program of China under Contracts No.
2020YFA0406300, No. 2020YFA0406400; National Natural Science Foundation of China (NSFC) under
Contracts No. 11635010, No. 11735014, No. 11835012, No. 11935015, No. 11935016, No. 11935018, No.
11961141012, No. 12005195, No. 12022510, No. 12025502, No. 12035009, No. 12035013, No. 12192260,
No. 12192261, No. 12192262, No. 12192263, No. 12192264, No. 12192265, No. 12225509; the Chinese
Academy of Sciences (CAS) Large-Scale Scientific Facility Program; Joint Large-Scale Scientific
Facility Funds of the NSFC and CAS under Contract No. U1832207; CAS Key Research Program of
Frontier Sciences under Contract No. QYZDJ-SSW-SLH040; 100 Talents Program of CAS; INPAC and
Shanghai Key Laboratory for Particle Physics and Cosmology; ERC under Contract No. 758462;
European Union's Horizon 2020 research and innovation program under Marie Sklodowska-Curie Grant
Agreement under Contract No. 894790; German Research Foundation DFG under Contracts No. 443159800,
Collaborative Research Center CRC 1044, GRK 2149; Istituto Nazionale di Fisica Nucleare, Italy;
Ministry of Development of Turkey under Contract No. DPT2006K-120470; National Science and
Technology fund; STFC (United Kingdom); The Royal Society, UK under Contracts No. DH140054,
No. DH160214; The Swedish Research Council; U. S. Department of Energy under Award No.
DE-FG02-05ER41374.

\clearpage

\textbf{Supplemental Material:A Brief Description of NREFT amplitude of $\eta'\to\eta\pi^0\pi^0$ Decay}

This supplemental material is based on Ref.~\cite{Kubis:2009sb}. In the $\eta'\to\eta\pi^0\pi^0$
decay
\begin{equation}
    \eta'(P_{\eta'})\to\pi^0(p_1)\pi^0(p_2)\eta(p_3),
\end{equation}
the kinematical variables $s_i$ are defined as $s_i=(P_{\eta'}-p_i)^2, i = 1,2,3$, and
$s_1 + s_2 + s_3 = M^{2}_{\eta'} + M^{2}_{\eta} + 2M^{2}_{\pi^0}$. The Dalitz plot distribution
of this decay can also described by kinematical variables $X$ and $Y$
\begin{equation}
    \begin{split}
        X&=\frac{\sqrt{3}|s_1-s_2|}{2M_{\eta'}Q_{\eta'}}=\frac{\sqrt{3}|T_{\pi^0_1}-T_{\pi^0_2}|}{Q_{\eta'}}, \\
        Y&=\frac{(M_{\eta}+2M_{\pi^0})[(M_{\eta'}-M_{\eta})^2-s_3]}{2M_{\eta'}M_{\pi^0}Q_{\eta'}} - 1 \\
         &=\frac{(M_{\eta}+2M_{\pi^0})T_{\eta}}{M_{\pi^0}Q_{\eta'}} - 1,
    \end{split}
\end{equation}
where $T_i$ denote kinetic energy of mesons in the rest frame of $\eta'$, and
$Q_{\eta'}=M_{\eta'}-M_{\eta}-2M_{\pi^0}$. The Dalitz plot distribution can be expanded by $X$ and
$Y$ around the center of the Dalitz plot
\begin{equation}
    |\mathcal{M}(X,Y)|^2=|\mathcal{N}|^2(1+aY+bY^2+cX+dX^2+\cdots),
\end{equation}
which is known as general parameterization. Here $\mathcal{N}$ is a normalization factor and
parameter $c$ is fixed at 0 since two $\pi^0$s are identical bosons. The general parameterization
can be also expressed as
\begin{equation}
    \mathcal{M}(X,Y)=\mathcal{N}\{1+\frac{a}{2}Y+\frac{1}{2}(b-\frac{a^2}{4})Y^2+\frac{d}{2}X^2+\cdots\}.
\end{equation}

The NREFT amplitude of $\eta'\to\eta\pi\pi$ can be decomposed to
\begin{equation}
    \begin{split}
        \mathcal{M}_{\eta'\to\eta\pi^0\pi^0}&=\mathcal{M}^{tree}_N+\mathcal{M}^{one-loop}_N+\mathcal{M}^{two-loop}_N+\cdots, \\
        \mathcal{M}_{\eta'\to\eta\pi^+\pi^-}&=\mathcal{M}^{tree}_C+\mathcal{M}^{one-loop}_C+\mathcal{M}^{two-loop}_C+\cdots.
    \end{split}
\end{equation}
The tree level amplitudes are
\begin{equation}
    \begin{split}
        \mathcal{M}^{tree}_N(s_1,s_2,s_3)&=\sum_{i=0}^{2}G_iX^i_3+G_3(X_1-X_2)^2, \\
        \mathcal{M}^{tree}_C(s_1,s_2,s_3)&=\sum_{i=0}^{2}H_iX^i_3+H_3(X_1-X_2)^2,
    \end{split}
\end{equation}
where $X_k=p^0_k-M_{\eta}, k=1,2,3$, $p^0_i$ is the energy of particle $i$ in the $\eta'$ rest
frame, and parameters $G_i$ are the low-energy coupling coefficients of
$\eta'\to\eta\pi^0\pi^0$ decay and $H_i$ for $\eta'\to\eta\pi^+\pi^-$ decay. The charged decay mode
is introduced for the further description of loop level amplitude and we assume $H_i=-G_i$
according to the isospin limit~\cite{Kubis:2009sb,Adlarson:2017wlz,BESIII:2017djm}. $G_i$ can be
evaulated by matching to the general parameterization
\begin{equation}
    \begin{split}
        G_0&=\mathcal{N}\{1-\frac{a}{2}+\frac{1}{2}(b-\frac{a^2}{4})\}, \\
        G_1&=\mathcal{N}\{\frac{a}{2}-(b-\frac{a^2}{4})\}\frac{M_{\eta}+2M_{\pi^0}}{M_{\pi^0}Q_{\eta'}}, \\
        G_2&=\mathcal{N}(b-\frac{a^2}{4})\frac{(M_{\eta}+2M_{\pi^0})^2}{2M^2_{\pi^0}Q^2_{\eta'}}, \\
        G_3&=\mathcal{N}\frac{3d}{2Q^2_{\eta'}}.
    \end{split}
\end{equation}

The loop level amplitude of $\eta'\to\eta\pi^0\pi^0$ decay are
\begin{equation}
    \begin{split}
        \mathcal{M}^{one-loop}_N(s_1,s_2,s_3)&=\mathcal{B}_{N1}(s_3)J_{+-}(s_3) \\
                                             &+\mathcal{B}_{N2}(s_3)J_{00}(s_3), \\
        \mathcal{M}^{two-loop}_N(s_1,s_2,s_3)&=C_{00}(s_3)\mathcal{B}_{N2}(s_3)J^2_{00}(s_3) \\
                                             &+C_{00}(s_3)\mathcal{B}_{N1}(s_3)J_{00}(s_3)J_{+-}(s_3) \\
                                             &+2C_{x}(s_3)\mathcal{B}_{C2}(s_3)J_{00}(s_3)J_{+-}(s_3) \\
                                             &+2C_{x}(s_3)\mathcal{B}_{C1}(s_3)J^2_{+-}(s_3),
    \end{split}
\end{equation}
with one-loop function
\begin{equation}
    \begin{split}
        J_{ab}&=\frac{iq_{ab}(s_k)}{8\pi\sqrt{s_k}}, \\
        q^2_{ab}(s)&=\frac{\lambda(s, M^2_a, M^2_b)}{4s}, \\
        \lambda(a,b,c)&=a^2+b^2+c^2-2(ab+ac+bc),
    \end{split}
\end{equation}
and neutral channel polynomials
\begin{equation}
    \begin{split}
        \mathcal{B}_{N1}(s_3)&=2C_{x}(s_3)[\sum_{i=0}^2H_iX^i_3+H_3\frac{4Q^2_3}{3s_3}q^2_{+-}(s_3)], \\
        \mathcal{B}_{N2}(s_3)&=C_{00}(s_3)[\sum_{i=0}^2G_iX^i_3+G_3\frac{4Q^2_3}{3s_3}q^2_{00}(s_3)],
    \end{split}
\end{equation}
and charged channel polynominals
\begin{equation}
    \begin{split}
        \mathcal{B}_{C1}(s_3)&=2C_{+-}(s_3)[\sum_{i=0}^2H_iX^i_3+H_3\frac{4Q^2_3}{3s_3}q^2_{+-}(s_3)], \\
        \mathcal{B}_{C2}(s_3)&=C_{x}(s_3)[\sum_{i=0}^2G_iX^i_3+G_3\frac{4Q^2_3}{3s_3}q^2_{00}(s_3)], \\
    \end{split}
\end{equation}
where
\begin{equation}
    \begin{split}
        C_{bc}(s_a)&=C_{bc}+4D_{bc}q^2_{bc}(s_a)+16F_{bc}q^4_{bc}(s_a), \\
        Q^2_a&=\frac{\lambda(M^2_{\eta'},M^2_a,s_a)}{4M^2_{\eta'}}.
    \end{split}
\end{equation}
The parameters $C_i$, $D_i$ and $F_i$ are coupling coefficients of $\pi\pi$ interaction and are
evaulated by matching to the effective range expansion of $\pi\pi$ scattering, where $i$
represent differnet $\pi\pi$ rescattering channels:
($00$)$\pi^0\pi^0\to\pi^0\pi^0$;($x$)$\pi^+\pi^-\to\pi^0\pi^0$;($+-$)$\pi^+\pi^-\to\pi^+\pi^-$,
\begin{equation}
    \begin{split}
        C_{00}&=\frac{16\pi}{3}(a_0+2a_2)(1-\xi), \\
        C_{x} &=\frac{16\pi}{3}(a_2-a_0)(1+\frac{\xi}{3}), \\
        C_{+-}&=\frac{8\pi}{3}(2a_0+a_2)(1+\xi), \\
        \xi   &=\frac{M^2_{\pi^{\pm}}-M^2_{\pi^0}}{M^2_{\pi^{\pm}}}, \\
        D_{00}&=\frac{4\pi}{3}(b_0+2b_2), \\
        D_{x} &=\frac{4\pi}{3}(b_2-b_0), \\
        D_{+-}&=\frac{2\pi}{3}(2b_0+b_2), \\
        F_{00}&=\frac{\pi}{3}(f_0+2f_2), \\
        F_{x} &=\frac{\pi}{3}(f_2-f_0), \\
        F_{+-}&=\frac{\pi}{6}(2f_0+f_2),
    \end{split}
\end{equation}
where $a_i$, $b_i$ and $f_i$ are S-wave scattering length, effective ranges and shape parameters of
isospin 0 and 2, respectively. $a_0$ and $a_2$ are taken as free or fixed parameters in differnet
cases in our study, and $b_i$ are fixed to theoretical value
$b_0=(0.276\pm0.006)\times M^{-2}_{\pi}$, $b_2=(-0.0803\pm0.0012)\times M^{-2}_{\pi}$, and $f_i$
are fixed to $0$. The $\pi\eta$ scattering terms are ignored because the $\pi\eta$ scattering is
much weaker than the $\pi\pi$ scattering.

\bibliography{ref.bib}

\end{document}